# Anisotropic magnetotransport in the layered antiferromagnet $TaFe_{1.25}Te_3$

Rajeswari Roy Chowdhury,[1 *] Samik DuttaGupta[2,3,4], Chandan Patra[1], Anshu Kataria[1], Shunsuke Fukami,[2,3,4,5,6] and Ravi Prakash Singh[1 †]

[1]*Department of Physics, Indian Institute of Science Education and Research Bhopal, Bhopal Bypass Road, Bhauri, Madhya Pradesh 462066. India*

[2]*Center for Science and Innovation in Spintronics, Tohoku University, 2-1-1 Katahira, Aoba-ku, Sendai 980-8577, Japan*

[3]*Laboratory for Nanoelectronics and Spintronics, Research Institute of Electrical Communication, Tohoku University, 2-1-1 Katahira, Aoba-ku, Sendai 980-8577, Japan*

[4]*Center for Innovative Integrated Electronic Systems, Tohoku University, 468-1 Aramaki Aza Aoba, Aoba-ku, Sendai 980-0845, Japan*

[5]*WPI-Advanced Institute for Materials Research, Tohoku University, 2-1-1 Katahira, Aoba-ku, Sendai 980-8577, Japan*

[6]*Inamori Research Institute for Science, Shijo, Shimogyo-ku, Kyoto 600-8411, Japan*

The discovery of fascinating ways to control and manipulate antiferromagnetic materials have garnered considerable attention as an attractive platform to explore novel spintronic phenomena and functionalities. Layered antiferromagnets (AFMs) exhibiting interesting magnetic structures, can serve as an attractive starting point to establish novel functionalities down to the two-dimensional limit. In this work, we explore the magnetoresistive properties of the spin-ladder AFM $TaFe_{1.25}Te_3$. Magnetization studies reveal an anisotropic magnetic behavior resulting in the stabilization of a spin-flop configuration for $H \perp$ (10-1) plane (*i.e.*, out-of-plane direction). Angle-dependent longitudinal and transverse magnetoresistances show an unusual anharmonic behavior. A significant anisotropic enhancement of magnetoresistance when $H \perp$ (10-1) plane compared to $H \parallel$ (10-1) directions has been observed. The present results deepen our understanding of the magnetoresistive properties of low-dimensional layered AFMs, and point towards the possibility of utilizing these novel material systems for antiferromagnetic spintronics.

## I. INTRODUCTION

The capability to manipulate and control antiferromagnets (AFMs) via electrical techniques has enunciated unprecedented opportunities for the development of materials, and new-concept electronic devices with novel functionalities [1-3]. Previous works utilizing AFMs in quasi three-dimensional heterostructure geometry have demonstrated promising characteristics such as electrical control of antiferromagnetic Néel vector and stabilization of AFM skyrmions [4-6], expected to pave the way for future AFM-based memories and devices for unconventional computing architectures. On the other hand, there has not been much focus on low-dimensional (layered and chain-like) AFMs, fundamentally and technologically interesting due to their abilities to introduce new functionalities and translate the existing properties down to the two-dimensional (2D) limit. Among the different families of layered AFMs, Fe-based chalcogenides are attractive owing to their reasonably high magnetic ordering temperature ($T_N$), coexistence and competing relationships between superconductivity and magnetism [8-12]. Amongst them, layered $TaFe_{1+x}Te_3$ is interesting [13,14], where the Fe atoms form a two-leg ladder along the principal axis (*i.e.*, *b*-axis) but with a zigzag shape representing an intriguing quasi-one-dimensional magnetic system. $TaFe_{1+x}Te_3$ crystallizes in a monoclinic structure, consisting of Ta-Fe bonded layers sandwiched between the Te layers (Fig. 1(a)). There are also excess Fe atoms, randomly occupying some interstitial sites, potentially influencing the magnetic order in the Fe-Te layers [15,16]. Previous magnetic and magnetoresistance measurements on $TaFe_{1.21}Te_3$ suggest a spin-density wave magnetic character, below $T_N$, where the neighboring spins within each ladder are antiparallelly coupled [17]. On the other hand, neutron diffraction measurements on $TaFe_{1.25}Te_3$ indicate a ferromagnetic two-leg zigzag ladder configuration which are antiferromagnetically coupled to their neighboring layers [18]. Despite these studies, a detailed investigation concerning the magnetoresistive responses originating from the unique ladder-like magnetic structure and its modification with varying number of layers remains unexplored. Owing to the layer dependent magnetic character, $TaFe_{1+x}Te_3$ serves as an archetype system for understanding the impact of dimensionality on magnetic order, electronic, topological, and correlated physics and tailoring of these properties for future development of novel layered materials. The interplay of this unique AFM order and their preferential alignment with respect to certain crystallographic directions can manifest in anisotropic magnetoresistive behavior, providing qualitative information concerning magnetic anisotropy, a key parameter concerning the potential of a material system for development of spintronic devices. Intuitively, the application of magnetic field (*H*) on this A-type interlayer antiferromagnetic coupling could also be interesting, where a potential transformation of the relatively weak AFM interaction into an Ising-like or XY-type ferromagnet (FM) state [19-22] can consequently result in a significant magnetoresistance (MR). While such an alteration of AFM order by *H* is generally deemed to be difficult, a strong coupling between the spin, and lattice degrees of freedom can result in a modification of electronic structure through a spin-flop transition, manifesting in significant MR. In addition, MR, and its angular dependence in naturally occurring layered compounds can also serve as an alternative route for understanding of the exotic magnetic structure. Finally, the existence of such a unique MR effect is expected to introduce new avenues for utilization of AFMs.

In this work, we have explored the magnetic, temperature and angle-dependent MR effects in layered AFM $TaFe_{1+x}Te_3$. Temperature dependent magnetization (*M*) measurements show antiferromagnetic ordering at ≈ 200 K, consistent with previous study [17]. An applied magnetic field ($H$) perpendicular to (10-1) plane show the existence of a spin-flop transition at $T \approx 130$ K, associated with an increase of $M$, and abrupt drop of the longitudinal MR, only for $H \perp$ (10-1) plane. Interestingly, the spin flop transition also results in a sharp deviation of the angle-dependent longitudinal MR behavior from its usual harmonic nature, manifesting in a strong anharmonicity in angular dependence. Along with this, there is a significant enhancement of longitudinal MR, compared to $H \parallel$ (10-1) configuration. Our work deepens the understanding of MR properties in layered AFMs and indicates the possibility of utilizing this magnetoresistive effect as a prospective scheme for introducing spintronic functionalities in layered AFMs.

[*]rajeswari@iiserb.ac.in
[†]rpsingh@iiserb.ac.in

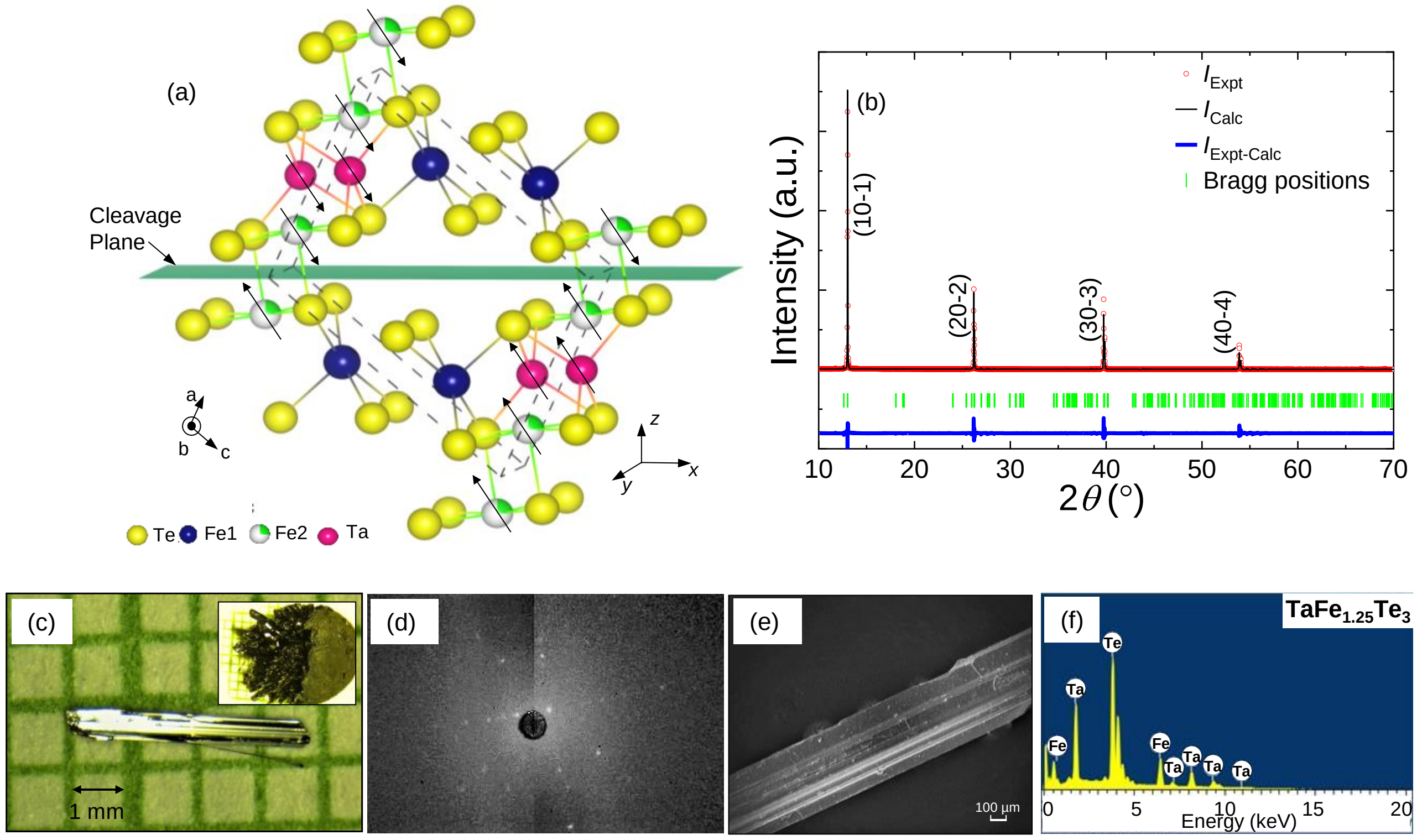

FIG. 1. (a) Crystal structure of $TaFe_{1+x}Te_3$. The interstitial Fe atoms are denoted by white-green circles. (b) Out-of-plane X-ray diffraction pattern with Le Bail fit for $TaFe_{1.25}Te_3$ single-crystalline sample at room temperature. (c) Optical micrograph of $TaFe_{1.25}Te_3$ single crystal utilized in this study. Inset shows the needle-like single crystals obtained by chemical vapor transport synthesis. (d) Laue diffraction pattern of the single crystal. (e) Scanning electron micrograph image of the needle-like nature of $TaFe_{1.25}Te_3$ single crystal. (f) Energy dispersive X-ray analysis (EDX) spectra for $TaFe_{1.25}Te_3$.

## II. EXPERIMENTAL DETAILS

Single crystals of $TaFe_{1.25}Te_3$ (TFT, hereafter) were grown by chemical vapor transport (CVT) method. A stoichiometric mixture of Ta (3N), Fe (3N) and Te (3N) were ground thoroughly and sealed in evacuated quartz tube along with $TeCl_4$ as transport agent. The tube was kept in a two-zone furnace at a temperature gradient of 690 °C/660 °C. Needle shaped crystals were obtained after ten days. Structural analysis of the crystals was performed by x-ray diffraction (XRD) at room temperature using a PANalytical diffractometer with Cu-$K_\alpha$ radiation. Magnetic properties were characterized using a superconducting quantum interference (SQUID) in the temperature range 5-300 K. Electrical and magnetotransport measurements were performed by a physical property measurement system using a conventional four-probe technique under an applied dc $I$ (|| (10-1) plane) of magnitude 10 mA. Longitudinal ($\rho_L$) and transverse ($\rho_H$) resistivities were obtained as a function of temperature ($T$) and applied magnetic field ($H$).

## III. RESULTS AND DISCUSSIONS

In this section, we show the experimental results and discuss the magnetic and magnetotransport properties of TFT single crystals.

### A. Structural characterization of $TaFe_{1.25}Te_3$ (TFT)

Figure 1(a) shows the crystal structure of TFT. The layered ternary compound crystallizes in a monoclinic structure (space group $P2_1$/m), comprising layers of Ta-Fe sandwiched between Te layers, forming $TaFeTe_3$. The excess Fe (atomic percentage 0.25) is expected to partially occupy the interstitial sites in a random manner (shown as white-green spheres in Fig. 1(a)). Figure 1(b) shows the experimental results of out-of-plane XRD for TFT single-crystals. The observed Bragg peaks can be indexed with ($l$0-$l$) peaks, perpendicular to the sample surface. Besides, any additional peak corresponding to unreacted elements or due to unintentional formation of other Ta-Fe-Te variants was not observed. Le Bail fit of the XRD data yields lattice parameters to be, a = 7.424 Å, b = 3.639 Å, c = 10.002 Å and $\beta$ = 109.132° respectively, consistent with earlier reports [13,14]. Figure 1(c) shows the optical microscope image of needle like single crystals of TFT. Inset shows the collection of as-grown crystals. Figure 1(d) shows the Laue diffraction pattern confirming formation of good quality of the obtained crystals. Scanning electron microscope (SEM) image of a needle like layered crystal is shown in Fig. 1(e). Chemical composition of the grown crystals was confirmed from atomic percentage ratios obtained from energy-dispersive X-ray (EDX) spectroscopy measurements (Fig. 1(f)) within the instrumental limit.

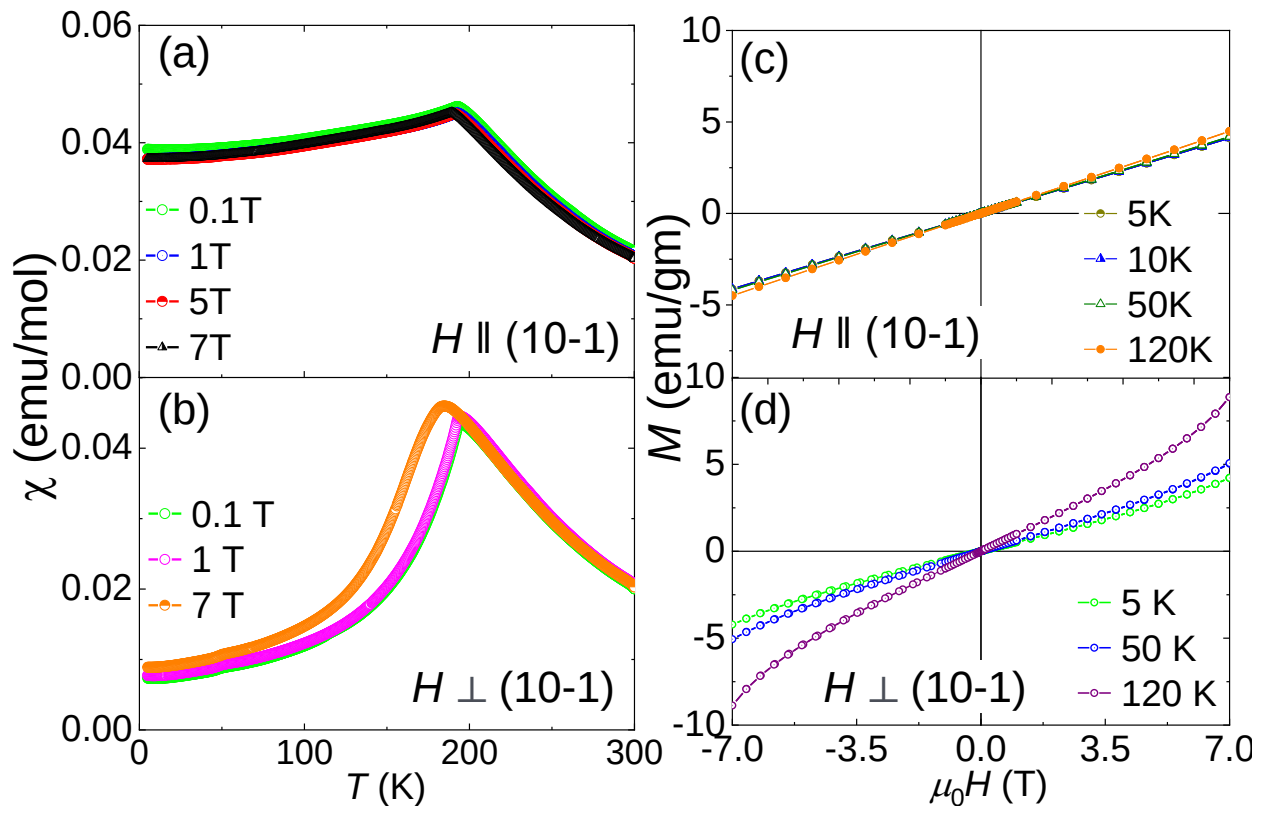

FIG. 2. (a) Magnetic susceptibility ($\chi$) versus temperature ($T$) under various magnetic fields $\mu_0 H$ = 0.1, 1, 5, and 7 T applied parallel to (10-1) plane, *i.e.*, along the plane of the crystal. (b) $\chi$ versus $T$ under various magnetic fields $\mu_0 H$ = 0.1, 1, and 7 T applied perpendicular to (10-1) plane, *i.e.*, along the out-of-plane direction. (c) Field ($H$) dependence of magnetization ($M$) for $TaFe_{1.25}Te_3$ single crystal at $T$ = 5, 10, 50, and 120 K, for applied $H$ parallel to (10-1) plane. (d) Experimental results for $M$ versus $H$ at $T$ = 5, 50, and 120 K, for applied $H$ perpendicular to (10-1) plane.

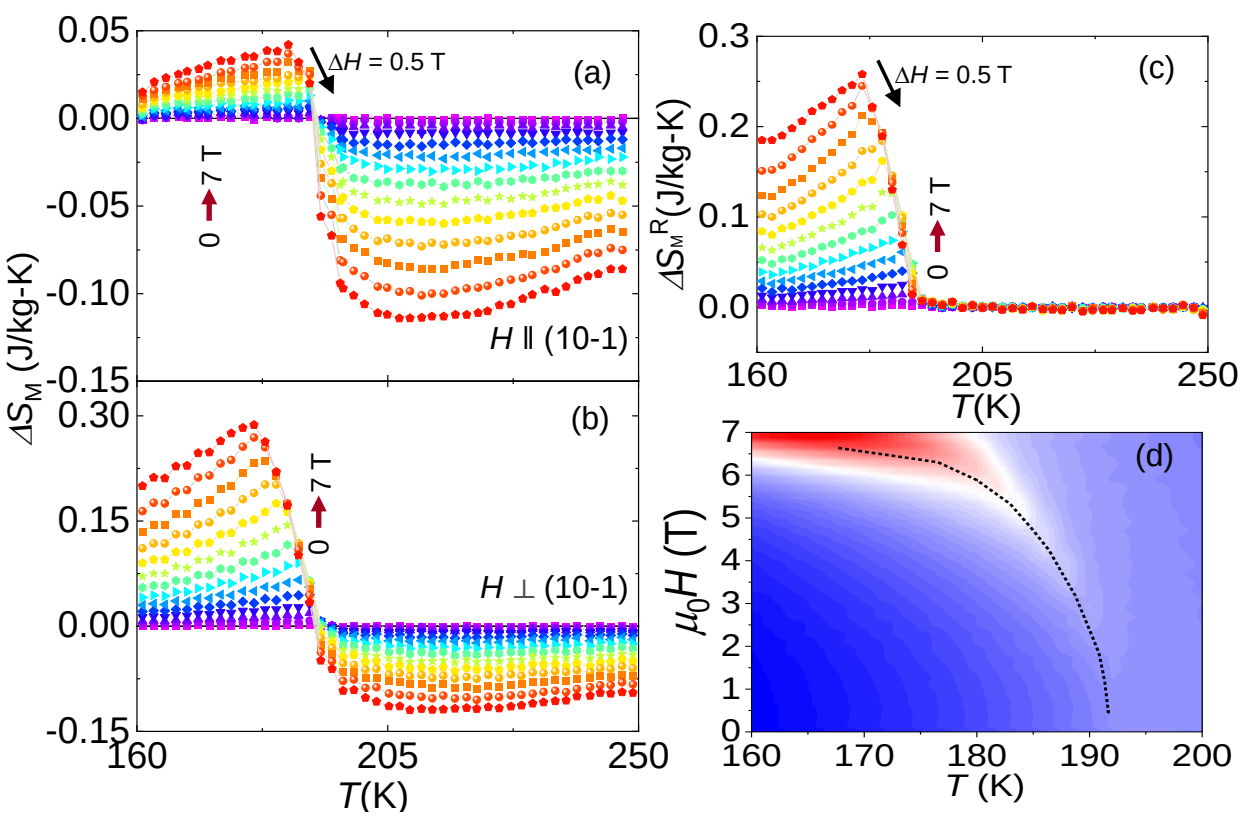

FIG. 3. (a) Temperature ($T$) dependence of magnetic entropy change ($\Delta S_M$) under application of magnetic fields $\mu_0 H$ = 1-7 T applied perpendicular to (10-1) plane. (b) Experimental results of similar measurements under application of magnetic fields $\mu_0 H$ = 1-7 T applied parallel to (10-1) plane. (c) $T$ dependence of rotational magnetic entropy change ($\Delta S_M^R$), defined as the difference between $\Delta S_M$ from (a), (b) at a constant $T$ and $H$. (d) $T$ dependence of spin flip field, obtained from $M$-$H$ measurements under applied $H \perp$ (10-1) plane.

### B. Magnetic and magnetocaloric properties of TFT

Figures 2(a), (b) show the experimental results of temperature ($T$) dependence of field-cooled (FC) magnetic susceptibility ($\chi$) under different magnetic field $\mu_0 H$ = 0.1, 1, 5, 7 T, applied parallel or perpendicular to the (10-1) sample plane (*i.e.*, in-plane or out-of-plane with respect to crystal), respectively. Magnetic susceptibility ($\chi$) under applied $\mu_0 H$ =0.1 T, along parallel or perpendicular to (10-1) plane shows a sharp transition at $T \approx 194$ K, demonstrating the onset of antiferromagnetic order. The observed Néel temperature ($T_N$) is slightly higher than previous report on single crystals [17,18], but almost matches with that of polycrystalline $TaFe_{1.25}Te_3$ ($T_N \approx 200$ K) [13]. When $H \perp$ (10-1) plane (i.e., along out-of-plane direction), $T_N$ is weakly suppressed from ≈ 194 K for $\mu_0 H$ = 0.1 T to ≈ 183 K for $\mu_0 H$ =7 T. Along with the reduction of $T_N$, some previous results also reported a ferromagnetic-like nature within a small temperature range below $T_N$, attributed to the alignment of excess Fe atoms towards the applied $H$ direction [23]. The coupling of this glass-like ferromagnetic state to the bulk AFM order was shown to result in a considerable exchange bias field of ≈ 0.16 T below 10 K [22]. Figures 2(c), (d) show the results of magnetization ($M$)-$H$ measurements for applied $H$ along (10-1) or $\perp$ (10-1) plane, respectively. For $5 < T < 50$ K, $M$ increases linearly with $H$, representative of an antiferromagnetic nature of the system. At higher $T$ ($\geq$ 120 K), for applied $H \perp$ (10-1) plane, we observe an increasing tendency above a threshold field, reminiscent of flopping of the antiferromagnetic moments leading to an enhancement of $M$. On the other hand, for $H \parallel$ (10-1) plane, we do not observe such behavior up to maximum applied $\mu_0 H$ = 7 T. Unlike the previous report [23], we also do not observe any hysteretic feature in $M$-$H$, indicating a virtually non-existent spin-glass like state. Besides, we observe a considerable difference in magnitude of $\chi$ for applied $H \parallel$ (10-1) or $\perp$ (10-1) plane, clearly indicating the existence of an antiferromagnetic anisotropy in the system.

To further characterize the anisotropic magnetic properties of TFT, isothermal magnetic entropy change ($\Delta S_M$) was calculated, under applied $H \parallel$ (10-1) and $\perp$ (10-1) plane directions within a $T$ range of 160 – 250 K from magnetization isotherms. $\Delta S_M$ was obtained from $M$-$H$ curves using Maxwell's relation [24,25]

$$\Delta S_M(T,H) = \int_0^H \left[\frac{\partial S(T,H)}{\partial H}\right]_T dH \qquad (1)$$

$$= \int_0^H \left[\frac{\partial M(T,H)}{\partial T}\right]_H dH. \qquad (2)$$

Figures 3(a), (b) show the $T$ dependence of $\Delta S_M$ under $\mu_0 H$ = 1-7 T, applied $\parallel$ or $\perp$ (10-1) plane, respectively. As we decrease $T$ from 250 K, $\Delta S_M$ is initially negative when the system is in paramagnetic state. Around $T \approx 194$ K, $\Delta S_M$ changes sign owing to the onset of AFM order, coinciding with $T_N$ obtained from magnetization measurements. Subsequently, the application of $H$ in the antiferromagnetic state results in adiabatic cooling owing to the enhancement of

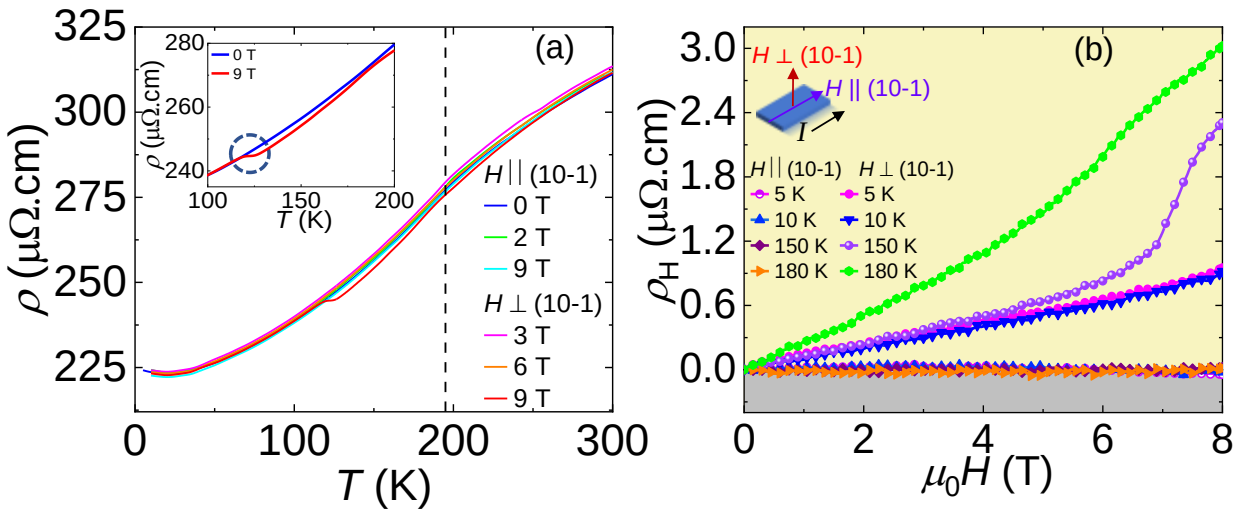

FIG. 4. (a) Temperature ($T$) dependence of longitudinal resistivity ($\rho_L$) under various magnetic field ($H$), applied parallel or perpendicular to (10-1) plane. Inset in (a) shows $\rho$ vs T under $\mu_0 H$ = 9 T is applied ⊥ (10-1) plane. Broken line indicates the $T$ at which there is a non-monotonic variation in $\rho$. (b) Applied $H$ dependence of transverse resistivity ($\rho_H$) at various $T$. Inset in (b) shows the direction of the applied $H$ and current ($I$) for $\rho$ and $\rho_H$ measurements.

configurational entropy of the spin structure, and reduction of the lattice entropy. Interestingly, under $\mu_0 H$ = 7 T, applied ⊥ (10-1) plane, $\Delta S_M$ = 0.29 J/kg-K, which is significantly larger than that for applied $H$ || (10-1) plane of identical magnitude ($\Delta S_M$ = 0.04 J/kg-K). We also calculate the rotational magnetic entropy change ($\Delta S_M^R$) as

$$\Delta S_M^R \ (T, H) = \Delta S_M\big(T, H_{\perp(10-1)}\big) - \Delta S_M\big(T, H_{||(10-1)}\big) \qquad (3)$$

Figure 3(c) shows the $T$ dependence of $\Delta S_M^R$ under applied $\mu_0 H$ = 1-7 T, confirming the anisotropic magnetic character of the system. The sign of $\Delta S_M^R$ (defined as Eq. (3)) is always positive, considered to be indirect evidence of strong AFM coupling among the Fe moments, perpendicular to the (10-1) plane. The magnitude of $\Delta S_M^R \approx$ 0.25 J/kg-K, for TFT, is comparable to van der Waals ferromagnetic $Fe_3GeTe_2$ [26]. Furthermore, we have observed that an increase of H results in a shift of the peak position towards lower $T$. The observed feature is most likely due to the flopping of the antiferromagnetic moments under applied $H$ which can be confirmed from Fig. 3(d), and, as shown later, plays a crucial role governing the unique magnetotransport features of this layered AFM system.

### B. Magnetotransport properties of $TaFe_{1.25}Te_3$ (TFT)

Figure 4(a) shows the $T$ dependence of longitudinal resistivity ($\rho_L$) under applied dc $I$ (|| (10-1) plane) *i.e.*, along crystal plane (inset of Fig. 4(b) shows the schematics of measurement set up). We observe a metallic behavior over the entire range with or without applied $H$ and a transition at $T \approx$ 195 K, approximately around the Néel temperature determined from magnetization measurement. When $\mu_0 H$ = 9 T is applied ⊥ (10-1) plane, we also observe an anomalous transition at $T \approx$ 120 K (inset of Fig. 4(a)), in agreement with the $M$-$H$ measurements (Fig. 2(d)). Figure 4(b) shows the $H$ dependence of transverse resistivity ($\rho_H$) at various $T$. Below T ≈ 150 K, $\rho_H$ increases linearly under $H$, applied either || or ⊥ (10-1) plane. Above this threshold, for $H$ ⊥ (10-1) plane, two distinct regimes are evident *viz.*, a linear regime followed by a sudden rise in $\rho_H$, tentatively attributed to arise from the net magnetization acquired by the flopping of the AFM moments. Carrier concentration calculated using the linear part of the field dependence of transverse resistivity below

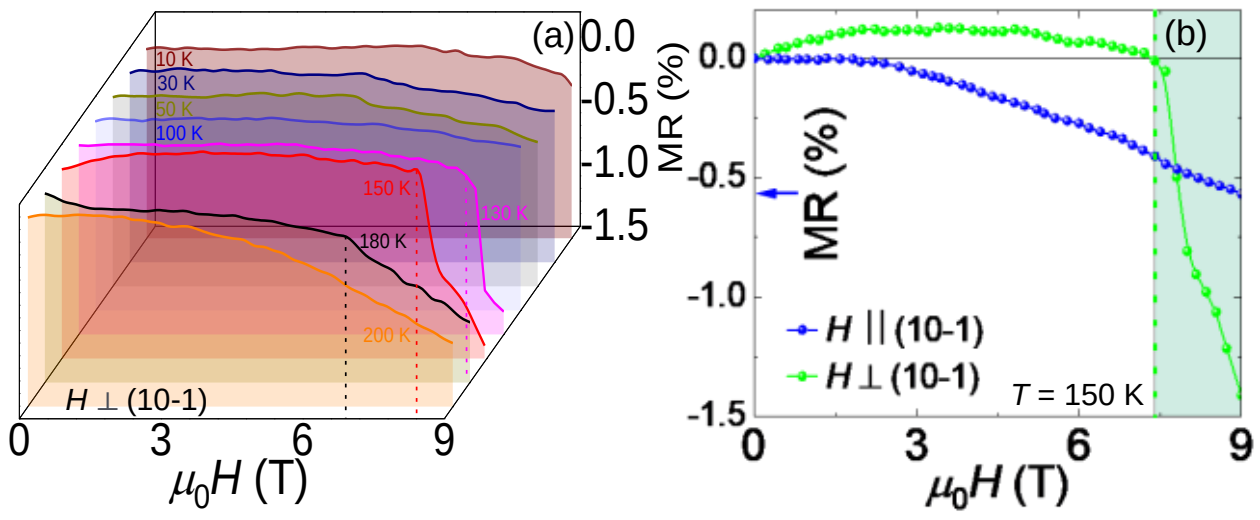

FIG. 5. (a) Temperature ($T$) dependence of magnetoresistance (MR) (in %) versus applied magnetic field ($H$), applied perpendicular to (10-1) plane (*i.e.*, $H$ || out-of-plane direction). Dotted arrows in (a) indicates the observed magnitude of $H$ at which the sharp drop in $\rho$ occurs, at $T$ = 130, 150 and 180 K. (b) Comparison of MR for applied $H$ || (10-1) or ⊥ (10-1) plane directions, at $T$ = 150 K.

spin flop field yields a carrier density of n ≈ 0.99 × $10^{21}$ $cm^{-3}$ at 150 K. To investigate the origin for the observed features, we measure $H$ dependence of $\rho_L$ at various $T$, and calculate magnetoresistance (MR) (in %) = $(\rho_L(H) - \rho_L(0))/\rho_L(0)$ (Fig. 5(a)). At low temperatures (10 ≤ $T$ ≤ 100 K), for both $H$ || or ⊥ (10-1) plane, MR exhibits a small negative amplitude (≤ 0.5%). However, an increase in $T$ ( ≥ 130 K), for $H$ ⊥ (10-1) plane, results in a slightly positive MR succeeded by a sharp drop. The threshold $H$ corresponding to this drop decreases from ≈ 8.4 T at $T$ = 130 K, to ≈ 6.4 T at $T$ = 180 K. Interestingly, no such behavior is observed for applied $H$ || (10-1) plane (Fig. 5(b)). To understand the factors responsible towards this distinct behavior, it is necessary to consider the impact of applied $H$ on the bulk antiferromagnetic order of TFT. TFT possess a zigzag ladder configuration of Fe moments and additional interstitial Fe moments, running parallel to the $b$-axis [15-18, 27]. However, they mainly differ on the magnetic arrangement of Fe moments within each zigzag ladder and their coupling to the subsequent layers. Some previous studies indicated intra-ladder antiferromagnetic arrangement of neighboring moments and spin-density wave type AFM structure [17], while neutron diffraction and angle-resolved photoemission spectroscopy measurements [18,27] indicate an intra-ladder ferromagnetic alignment of the Fe moments with an antiferromagnetic coupling between consecutive ones (A-type AFM order). Furthermore, this novel AFM ground state is also composed of a quasi-two-dimensional Fermi surface with sizeable inter-ladder hopping mediated by super exchange interaction [27]. The Fe moments within a ladder are arranged ferromagnetically at an angle of 17.6° with respect to the (10-1) direction (*i.e.*, the conventional out-of-plane direction), and are antiferromagnetically coupled to the nearest neighboring chains. Considering this A-type AFM ground state, the sharp drop in longitudinal MR for $H$ ⊥ (10-1) plane (Fig. 5(a)) might be attributed to arise from an inter-ladder spin-flop configuration and/or spin-flop followed by rotation of the AFM moments perpendicular to the applied $H$. The absence of a similar sharp drop in longitudinal MR for $H$ || (10-1) plane (Fig. 5(b)) indicates the existence of an anisotropic behavior, restricting the stabilization of the spin-flop configuration only for certain directions of applied $H$. Transverse resistivity ($\rho_H$) measurements also support a similar scenario (Fig. 4(b)), where the sudden rise in $\rho_H$ for $T \ge 150$ K can be attributed to arise from the net magnetization acquired by the spin-flopped configuration, for applied $H \perp$ (10-1) plane. As shown below, the stabilization of a unique AFM ground state, and the realization of an anisotropic spin-flop configuration renders drastic ramifications into the angular magnetoresistive properties of TFT.

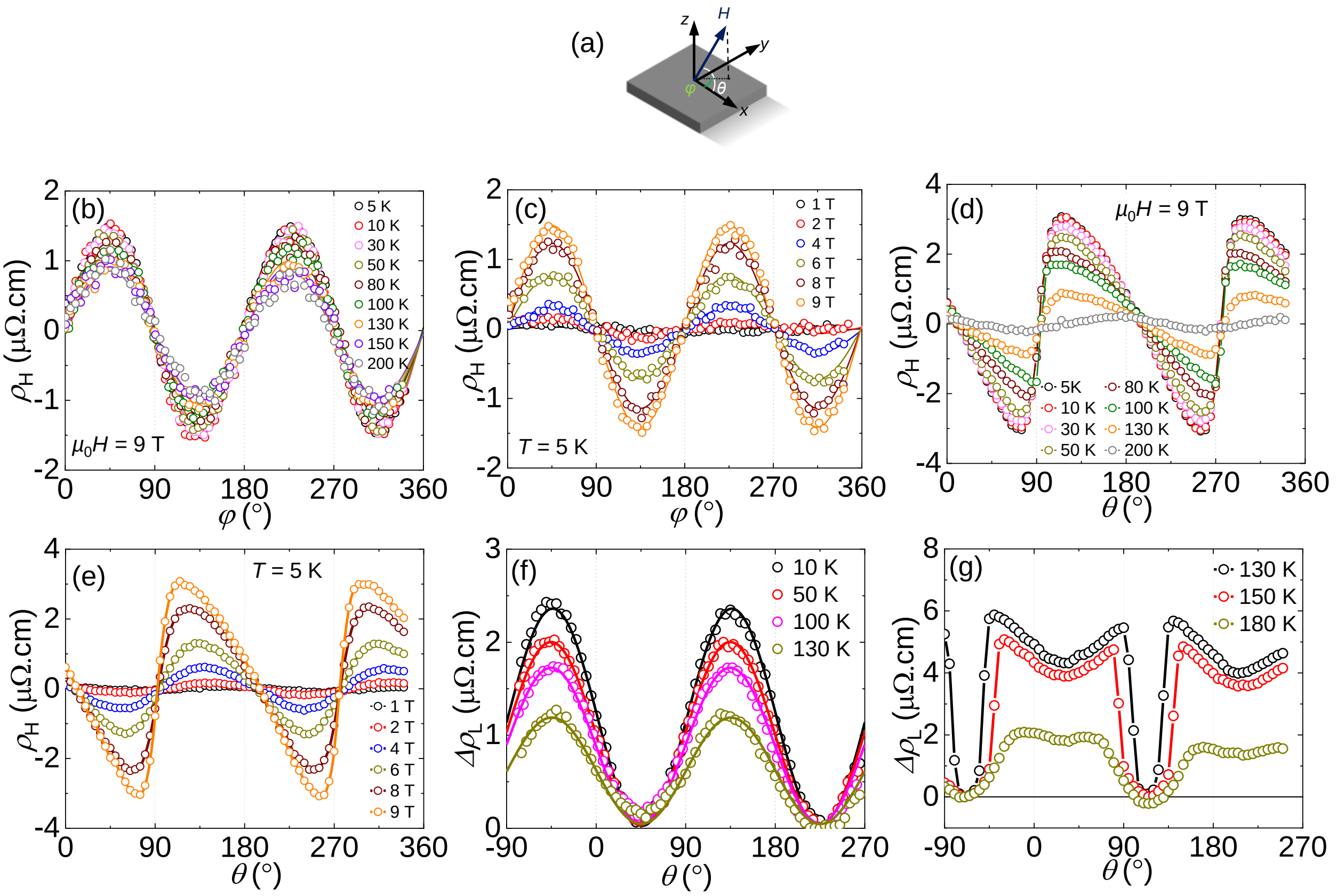

FIG. 6. (a) Schematic representation of the measurement configuration utilized in this work. The azimuthal ($\theta$) and polar ($\varphi$) angles are defined as the angle subtended by the applied magnetic field ($H$) with respect to the applied current ($I$) along $x$-direction. (b) $\varphi$ dependence of transverse resistivity ($\rho_H$) under constant magnetic field $\mu_0 H = 9$ T ($H \parallel$ (10-1) plane), at various temperatures ($T$). (c) $\varphi$ dependence of $\rho_H$ under varying magnitudes of $H$ at T = 5 K. (d) $\theta$ dependence of $\rho_H$ under constant magnetic field $\mu_0 H = 9$ T ($H \perp$ (10-1) plane), at various $T$. (e) $\theta$ dependence of $\rho_H$ under varying magnitudes of $H$, at T = 5 K. (f) $\theta$ dependence of longitudinal resistivity ($\rho_L$) under constant magnetic field $\mu_0 H = 6$ T at various $T$ ($10 \le T \le 130$ K). (g) Experimental results of similar measurements under $\mu_0 H = 8$ T, for $T \ge 130$ K. Solid lines in (b),(c), and (f) denotes the fitting of the experimental data with the harmonic sine squared dependence.

To clarify the manifestations of the anisotropic magnetotransport behavior, we measured the angle dependent longitudinal and transverse magnetoresistance at various $T$. Figure 6(a) shows the schematic diagram of the measurement geometry, and the definitions of the azimuthal (angle $\varphi$ between $H$ and $I$) and polar (angle $\theta$ between $H$ and $I$) angles with respect to the single crystal. An applied dc $I$ ($\parallel$ (10-1) plane) was passed through the single-crystal along the $x$-direction (*i.e.*, along crystal plane). The resulting change in longitudinal or transverse voltages were measured under rotation of external $H$ of constant magnitude along the azimuthal and polar planes. Intuitively, $\theta$ sweep corresponds to $H$ rotating from a magnetic hard ($\theta \approx 0°$) to easy ($\theta \approx 72.4°$) directions, compared to $\varphi$ sweep, enabling us to quantitatively clarify its impact on the magnetoresistive features. Figures 6(b), (c) show experimental results of $\rho_H$ versus $\varphi$, under constant $H$ ($T = 5$ K) and $T$ ($\mu_0 H = 9$ T), respectively. We observe that $\rho_H$ versus $\varphi$ follows a two-fold symmetry with a $\mathrm{Sin}^2\,\varphi$ behavior (solid lines in Figs. 6(b),(c)). Furthermore, the amplitude of the $\mathrm{Sin}^2\,\varphi$ dependence monotonically decreases with increasing $T$ or decreasing $H$. Owing to the A-type AFM order along with the absence of spin-flop configuration for $H \parallel$ (10-1) plane, the observed magnetoresistive behavior can be interpreted as anisotropic magnetoresistance (AMR), arising from the orientation of the antiferromagnetic Néel vector with respect to $I$, commonly found in most FM or AFMs. On the other hand, significant deviations from the conventional AMR [28,29] were observed for $H$ rotations along $\theta$ (Figs. 6(d), (e)). $\rho_H$ versus $\theta$ curves deviate significantly from the harmonic behavior and shows a sharp sign reversal as $H \perp$ (10-1) plane. Interestingly, the anharmonic behavior is strongly dependent on $H$ (Fig. 6(e)), and weakly varies with $T$ (Fig. 6(d)). This indicates that the underlying origin of this anharmonic behavior might not be dominantly linked to the flopping of the AFM moments, induced by the applied $H$. While we cannot rule out any contribution from spin-flop configurations, we speculate that the anharmonic nature might be strongly influenced by the magnetic anisotropy of TFT, preferring an AFM alignment perpendicular to the (10-1) planes. To further explore the peculiar nature of the magnetotransport features of TFT, we also study the effect of $H$ rotation ($\theta$) on $\rho_L$ (Figs. 6(f),(g)). For applied $H$ less than the spin-flop field ($10 \leq T \leq 130$ K, $\mu_0H$ = 6 T), the MR curves show a harmonic behavior ($\mathrm{Sin}^2\,\theta$ dependence). Interestingly, as $H$ is larger than the spin-flop field ($T \geq 130$ K, $\mu_0H$ =8 T) we observe a significantly different behavior, strongly anharmonic in nature with the appearance of plateau-like features for certain $\theta$, along with a significant increase in the magnitude of $\rho_L$. Furthermore, an increase in $T$, in this regime at fixed $\mu_0H$ = 8 T also results in a reduction of $\rho_L$ magnitude, attributed to an enhancement of spin fluctuation near $T_N$.

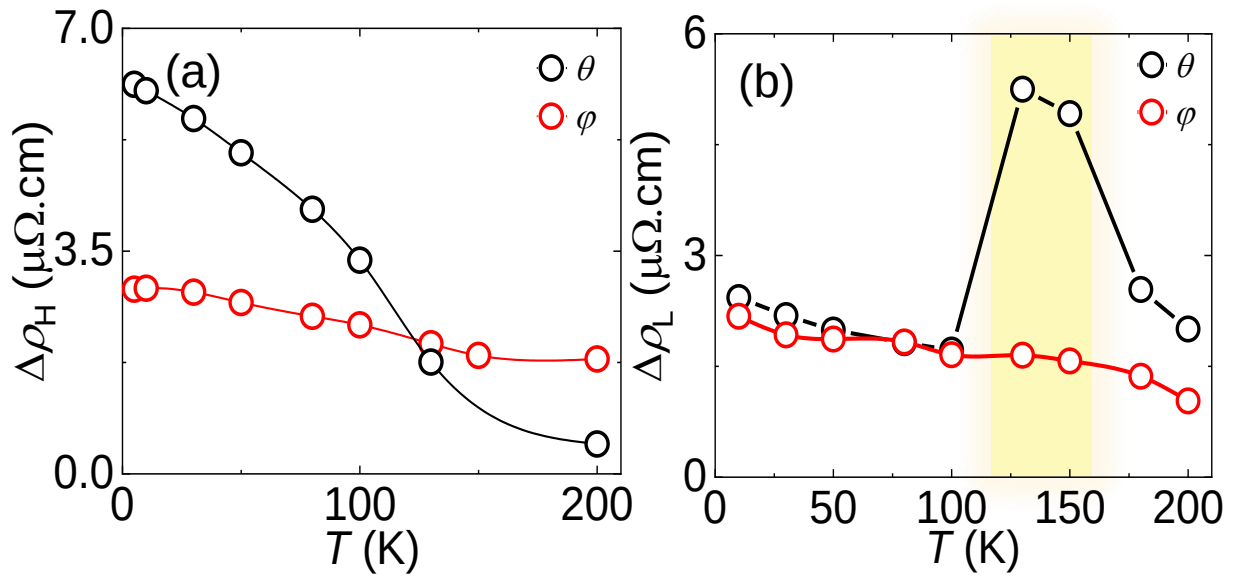

FIG. 7. (a) Temperature ($T$) dependence of the amplitude of transverse resistivity ($\rho_H$) for $H$ rotations along azimuthal ($\theta$) and polar ($\varphi$) directions under applied $\mu_0H$ = 8 T. (b) T dependence of the amplitude of longitudinal resistivity ($\rho_L$) for $H$ rotations along azimuthal ($\theta$) and polar ($\varphi$) directions under applied $\mu_0H$ = 8 T.

To get a deeper understanding of the underlying physics governing the different magnetoresistive behaviors, we have extracted the magnitude of the observed magnetoresistances versus $\theta$ and $\varphi$ (Figs. 7(a),(b)). For either $\theta$ or $\varphi$ rotations, the amplitude of the transverse resistivity ($\Delta\rho_H$) monotonically decreases with increasing $T$, as expected for a MR signal originating from magnetic order. On the other hand, the amplitude of longitudinal resistivity ($\Delta\rho_L$) shows a dramatic behavior. Starting from $T$ = 10 K, for $\theta$ or $\varphi$ rotations, $\Delta\rho_L$ slightly decreases up to 100 K, thereafter, showing a significant enhancement in the $T$ range 130-150 K (for $\theta$ rotation), while decreases monotonically for $\varphi$ rotations. Considering an A-type AFM ground state of TFT with an intra-layer ferromagnetic alignment, for $\varphi$ rotation, the $H$ is always perpendicular to the Néel vector, roughly oriented at ≈ 18° with respect to the surface normal. We speculate that no further discernable reorientation of the Néel vector occurs under $\varphi$ rotation resulting in a monotonic decrease of MR behavior. As stated before, $\theta$ rotation results in the $H$ rotating between magnetically easy and hard directions which can have a profound effect on the interlayer AFM moments. Interestingly, the observed sharp drop in $\rho_L$ occurs when $H$ is perpendicular to the (10-1) plane, *i.e.*, applied almost along the magnetically easy directions. As a result, the significant enhancement in $\Delta\rho_L$ is tentatively attributed to the stabilization of a spin-flop configuration of the interlayer AFM configuration. Our results are also indicative of a strong inter-layer magnetic coupling and represents an intriguing situation despite the large separation between the Fe moments between the adjacent layers. In a typical AFM, the magnitude of AMR originating either from the spin-orbit coupling of the electronic band structure or exchange coupling to an adjacent FM is small, roughly 0.5-1 % [28,29]. On the other hand, the MR effect in the spin-flop configuration is substantially large ≈ 3-4 %. We believe that this work would open an unexplored pathway to utilize spin-flop configurations for introducing novel functionalities in antiferromagnetic spintronics.

## IV. CONCLUSIONS

In conclusion, we have investigated the magnetic, temperature and angle-dependent longitudinal and transverse magnetoresistive effects in the layered AFM $TaFe_{1.25}Te_3$. Temperature and applied magnetic field dependent magnetization measurements reveal the existence of an anisotropic behavior, $H \perp$ (10-1) plane results in a spin flop-like transition around $T \geq 130$ K as opposed to an antiferromagnetic nature for $H \parallel$ (10-1) plane. The spin-flop behavior also manifests in anisotropic magnetotransport behavior and results in a sharp drop of linear resistivity at similar $T$ and $H$ configurations. Interestingly, the angle dependent longitudinal magnetoresistance around the spin-flop transition shows a strong anharmonic behavior along with a pronounced enhancement in its magnitude. We believe that our results will inspire future experimental investigations concerning modification of this unusual longitudinal MR around the spin-flop transition with the variation of number of layers and possible existence of unconventional spin textures to initiate a new paradigm of AFM spintronics with layered material systems. The present experimental results provide considerable insights into the remarkable magnetic, and magnetotransport feature of layered AFMs, and suggests an alternative scheme to introduce novel spintronic functionalities in layered AFMs.

## V. ACKNOWLEDGEMENTS

We thank H. Ohno for fruitful discussions. R. R. C. acknowledges Department of Science and Technology (DST), Government of India, for financial support (Grant no. DST/INSPIRE/04/2018/001755). R. P. S. acknowledges Science and Engineering Research Board (SERB), Government of India, for Core Research Grant CRG/2019/001028. A portion of this work was supported by JSPS Kakenhi 19H05622, 20K15155 and RIEC International Cooperative Research Projects, Tohoku University.

## REFERENCES

[1] V. Baltz, A. Manchon, M. Tsoi, T. Moriyama, T. Ono, and Y. Tserkovnyak, Rev. Mod. Phys. **90**, 015005 (2018).

[2] T. Jungwirth, X. Marti, P. Wadley, and J. Wunderlich, Nature Nanotech. **11**, 231 (2016).

[3] J. Železný, P. Wadley, K. Olejník, A. Hoffmann, and H. Ohno, Nature Phys. **14**, 220 (2018).

[4] P. Wadley, B. Howells, J. Železný, C. Andrews, V. Hills, R. P. Campion, V. Novák, K. Olejník, F. Maccherozzi, S. S. Dhesi, S. Y. Martin, T. Wagner, J. Wunderlich, F. Freimuth, Y. Mokrousov, J. Kuneš, J. S. Chauhan, M. J. Grzybowski, A. W. Rushforth, K. W. Edmonds, B. L. Gallagher, and T. Jungwirth, Science **351**, 587 (2016).

[5] S. Y. Bodnar, L. Šmejkal, I. Turek, O. Gomonay, J. Sinova, A. A. Sapozhnik, H.-J. Elmers, M. Kläui, and M. Jourdan, Nature Commun. **9**, 348 (2018)

[6] S. DuttaGupta, A. Kurenkov, O. A. Tretiakov, G. Krishnaswamy, G. Sala, V. Krizakova, F. Machherozzi, S. S. Dhesi, P. Gambardella, S. Fukami, and H. Ohno, Nature Commun. **11**, 5715 (2020).

[7] W. Legrand, D. Maccariello, F. Ajejas, S. Collin, A. Vecchiola, K. Bouzehouane, N. Reyren, V. Cros, and A. Fert, Nature Mater. **19**, 34 (2020).

[8] F. C. Hsu, J. Y. Luo, K. W. Yeh, T. K. Chen, T. W. Huang, P. M. Wu, Y. C. Lee, Y. L. Huang, Y. Y. Chu, D. C. Yan, and M. K. Wu, Proc. Natl. Acad. Sci. **105**, 14262 (2008).

[9] K. W. Yeh, T. W. Huang, Y. L. Huang, T. K. Chen, F. C. Hsu, P. M. Wu, Y. C. Lee, Y. Y. Chu, C. L. Chen, J. Y. Luo, D. C. Yan, and M. K. Wu, Europhys. Lett. **84**, 37002 (2008).

[10] M. H. Fang, H. M. Pham, B. Qian, T. J. Liu, E. K. Vehstedt, Y. Liu, L. Spinu, and Z. Q. Mao, Phys. Rev. B 78, 224503 (2008).

[11] M. Bendele, A. Amato, K. Conder, M. Elender, H. Keller, H. H. Klauss, H. Luetkens, E. Pomjakushina,A.Raselli, andR.Khasanov, Phys. Rev. Lett. 104, 087003 (2010).

[12] S. Margadonna, Y. Takabayashi, Y. Ohishi, Y. Mizuguchi, Y. Takano, T. Kagayama, T. Nakagawa, M. Takata, and K. Prassides, Phys. Rev. B 80, 064506 (2009).

[13] M. E. Badding, J. Li, F. J. DiSalvo, W. Zhou, and P. P. Edwards, J. Solid State Chem. 100, 313 (1992)

[14] C. Pérez Vicente, M. Womes, J. C. Jumas, L. Sánchez, and J. L. Tirado, J. Phys. Chem. B. 102, 8712 (1998).

[15] S. Li, C. de la Cruz, Q. Huang, Y. Chen, J. W. Lynn, J. Hu, Y.-L. Huang, F.-C. Hsu, K.-W. Yeh, and P. Dai, Phys. Rev. B. 79, 054503 (2009).

[16] L. Zhang, D. J. Singh, and M. H. Du, Phys. Rev. B. 79, 012506 (2009).

[17] R. H. Liu, M. Zhang, P. Cheng, Y. J. Yan, Z. J. Xiang, J. J. Ying, X. F. Wang, A. F. Wang, G. J. Ye, X. G. Luo, and X. H. Chen, Phys. Rev. B. 84, 184432 (2011)

[18] X. Ke, B. Qian, H. Cao, J. Hu, G. C. Wang, and Z. Q. Mao, Phys. Rev. B. 85, 214404 (2012)

[19] Y. Ando, A. N. Lavrov, and S. Komiya, Phys. Rev. Lett. 90, 247003 (2003).

[20] A. N. Lavrov, H. J. Kang, Y. Kurita, T. Suzuki, S. Komiya, J. W. Lynn, S. H. Lee, P. C. Dai, and Y. Ando, Phys. Rev. Lett. 92, 227003 (2004).

[21] J. Wang, J. Deng, X. Liang, G. Gao, T. Ying, S. Tian, H. Lei, Y. Song, X. Chen, J. Guo, and X. Chen, Phys. Rev. Mater. 5, L091401 (2021).

[22] H. Wang, C. Lu, J. Chen, Y. Liu, S. L. Yuan, S.-W. Cheong, S. Dong, and J.-M. Liu, Nature Commun. **10**, 2280 (2019).

[23] Y. Liu, J. J. Bao, C. Q. Xu, W. H. Jiao, H. Zhang, L. C. Xu, Z. Zhu, H. Y. Yang, Y. Zhou, Z. Ren, P. K. Biswas, S. K. Ghosh, Z. Yang, X. Ke, G. H. Cao, and X. Xu, Phys. Rev. B. **104**, 104418 (2021).

[24] T. Krenke, E. Duman, M. Acet, E. F. Wassermann, X. Moya, L. Mañosa, and A. Planes, Nature Mater. **4**, 450 (2005).

[25] R. Roy Chowdhury, S. Dhara, and B. Bandyopadhyay, Physica B **517**, 6 (2017).

[26] Y. Liu, J. Li, J. Tao, Y. Zhu, and C. Petrovic, Sci. Rep. **9**, 13233 (2019).

[27] X. Min, W. L.-Min, P. Rui, G. Q.-Qin, C. Fei, Y. Z.-Rong, Z. Yan, C. S.-Di, X. Miao, L. R.-Hua, S. Ming, C. X.-Hui, Y. W.-Guo, K. Wei, X. B.-Ping, F. D.-Lai, M. Arita, K. Shimada, H. Namatame, M. Taniguchi, M. Matsunami, and S. Kimura, Chinese Phys. Lett. **32**, 027401 (2015).

[28] X. Marti, I. Fina, C. Frontera, J. Liu, P. Wadley, Q. He, R. J. Paull, J. D. Clarkson, J. Kudrnovský, I. Turek, J. Kuneš, D. Yi, J.-H. Chu, C. T. Nelson, L. You, E. Arenholz, S. Salahuddin, J. Fontcuberta, T. Jungwirth, and R. Ramesh, Nature Mater. **13**, 367 (2014).

[29] I. Fina, X. Marti, D. Yi, J. Liu, J. H. Chu, C. R.-Serrao, S. Suresha, A. B. Shick, J. Železný, T. JUngwirth, J. Fontcuberta, and R. Ramesh, Nature Commun. **5**, 4671 (2014).